# Enabling automatic transcription of child-centered audio recordings from real-world environments

Daniil Kocharov (ORCID ID: 0000-0002-7858-5331, daniil.kocharov@tuni.fi)[1], Azarias Galama (azarias.galama@tuni.fi)[1], Okko Räsänen (ORCID ID: 0000-0002-0537-0946, okko.rasanen@tuni.fi)[1]

[1] Signal Processing Research Centre, Tampere University, Finland,
Faculty of Information Technology and Communication Sciences,
Department of Computing Sciences,
P.O. Box 553, FI-33101 Tampere, Finland

## Abstract

Longform audio recordings obtained with microphones worn by children—also known as child-centered daylong recordings—have become a standard method for studying children's language experiences and their impact on subsequent language development. Transcripts of longform speech audio would enable rich analyses at various linguistic levels, yet the massive scale of typical longform corpora prohibits comprehensive manual annotation. Meanwhile, automatic speech recognition (ASR)-based transcription faces significant challenges due to the noisy, unconstrained nature of real-world audio. Previous attempts have assumed that ASR must process each longform recording in its entirety. In this work, we present an approach to automatically detect those utterances in longform audio that can be reliably transcribed with modern ASR systems, allowing automatic and relatively accurate transcription of a notable proportion of all speech in typical longform data. We validate the approach on four English longform corpora, showing that it achieves a median word error rate (WER) of 0% and a mean WER of 16% when transcribing 30% of the total speech in the dataset. In contrast, transcribing all speech without any filtering yields a median WER of 52% and a mean WER of 51%. We also compare word log-frequencies derived from the automatic transcripts with those from manual annotations and show that the frequencies correlate at $r$ = 0.94 (Pearson) for all transcribed words and $r$ = 0.99 for words that appear at least five times in the automatic transcripts. Overall, the work provides a concrete step toward increasingly detailed automated linguistic analyses of child-centered longform audio.

## 1. Introduction

In order to scientifically study and understand child language development, it is crucial to understand the language input that children hear in their everyday lives. A standard modern practice for studying children's language experiences is to use wearable recorders, such as the LENA system (Xu et al., 2008; Gilkerson & Richards, 2009), which a child wears throughout an entire day, thereby capturing all audio in the vicinity of the child in addition to the child's own vocalizations. By placing the recorder into a special pocket sewn into a wearable garment, the same setup can be used to study the sound environments of infants, toddlers, and older children alike. Due to the non-invasive nature of the setup and the consequent caregiver attunement to being recorded (e.g., Scaff et al., 2023), the recordings can be considered much more representative of children's language input—that is, more naturalistic—compared to audio recordings conducted in laboratory settings (e.g., during structured or free play) or audio tracks extracted from at-home video recordings (Bergelson et al., 2019).

However, the naturalistic and thereby unconstrained audio recordings come at a cost: the audio quality of speech in the recordings is extremely variable and generally much worse than what is typical of studio recordings, near-field conditions (mobile phones, headsets with lavalier microphones, etc.) or other controlled setups (e.g., crowdsourced speech recordings; Kearns, 2014). For instance, Räsänen et al. (2019) and Lavechin et al. (2024) estimate around 0–10 dB average speech-to-noise ratio (SNR) for child-centered longform audio with high variability from completely unintelligible speech to occasional high SNR audio. Moreover, typical child-directed speech (CDS) can be quite distinct from everyday adult-to-adult conversational or formal language, including short utterances, exaggerated prosody, atypical lexical items and grammatical constructs, and generally highly situated communication with a broad range of communicative intents across various contexts (see, e.g., Nikolaus et al., 2022), which also shift in distribution as children grow up. Together, these factors (audio quality, atypical language) make automatic speech recognition (ASR) very challenging on longform data, even though ASR transcripts would be extremely useful for developmental research.

In fact, ASR has been found to be so challenging on longform audio that the current state-of-the-art approach is to simply count the number of words or other linguistic units in the speech heard by the child, as implemented in the LENA analysis software and its open-source alternative ALICE (Räsänen et al., 2021)[1]. At the same time, several attempts have been made in different labs around the world (including ours) to apply state-of-the-art ASR for longform audio, such as the Whisper ASR system by OpenAI (Radford et al., 2023) (personal communication). However, we are not aware of successful solutions to the problem, as the existing attempts have resulted in unusable transcriptions due to very high word error rates (WER) (Tang et al., 2025). This is despite ASR nowadays being largely considered a 'solved problem' for more controlled signal conditions.

[1] Another consideration in favor of "unit counting" is greater language independence of the automatic processing algorithms compared to resource-intense ASR, enabling comparable analyses both on high- and low-resource languages.

The lack of automatic transcription for longform audio poses a critical limitation to the research that can be conducted with child-centered audio. Longform audio corpora can comprise thousands of hours of audio (e.g., the SEEDLingS corpus; Bergelson et al., 2019), which means that the audio cannot be comprehensively manually transcribed for content due to the sheer scale of the data. Without transcriptions, it is impossible to study linguistic characteristics of child language input in detail, such as analyzing the relationship between heard vs. learned lexical items, studying CDS syntax and its change as a function of child development, or quantifying individual variability in language experiences in terms of linguistic content. While manual annotations targeted at small subsets of data or specific aspects of language can be used in research (e.g., Bergelson et al., 2019), they are still sparse compared to overall audio duration despite the large human efforts required for manual transcription (see, e.g., Casillas, 2022), preventing systematic and representative (dense) analyses of the input. This calls for better solutions to automatically transcribe child-centered longform recordings.

In this paper, we address the need for longform ASR and describe a technical advance that enables automatic transcription of longform speech contents. Importantly, we do not pursue error-free ASR for *all speech* in longform audio, but we build on an assumption that *a significant proportion* of the speech in a recording could be transcribed accurately due to sufficient signal quality of the corresponding audio segments. We describe a method for automatic detection of the segments of audio on which ASR output can be trusted, thereby enabling automatic transcription of speech in longform audio at a scale at least an order of magnitude greater than what has been achieved with manual efforts.

We validate our approach with audio recordings and manual transcriptions from four longform child-centered corpora, reporting overall WERs and comparing word frequency estimates from manual versus automatic transcriptions. As a result, we show how the pipeline produces usable fully automatic estimates of speech content with low WERs. For practical reasons, we limit the current work to different varieties of English, but the general methodology should be applicable to other languages with sufficiently powerful ASR models available to them. We also limit our analysis to adult speech only, as the automatic recognition of children's speech poses yet another major challenge for current ASR technologies, even in the case of high-quality audio recordings (Patel & Scharenborg, 2024; Singh et al., 2026).

## 2. Data and Methods

### 2.1 Data

To develop and evaluate our system, we used audio data from four child-centered longform audio corpora with manual annotations. These comprise English corpora from the Analyzing Child Language

Experiences around the World (ACLEW) project[2], i.e., manually annotated parts of the SEEDLingS corpus (Bergelson, 2017), the LuCiD Language 0–5 corpus (Rowland et al., 2018), the McDivitt/Winnipeg corpus (McDivitt & Soderstrom, 2016), and the San Joaquin Valley Corpus (Warlaumont et al., 2016).

The speech data are very diverse in terms of both signal quality and speech content, as all the data consist of unconstrained recordings from everyday environments of young children collected with LENA recorders. The SEEDLingS corpus contains 522 daylong recordings from 44 US English-speaking families in New York State recorded monthly when infants were 6 to 18 months of age (Bergelson, 2017; Bergelson et al., 2019). The LuCiD Language 0–5 corpus contains 235 daylong recordings from 39 UK English-speaking families from Northwest England with infant ages ranging from 11 to 39 months (Rowland et al., 2018). The McDivitt/Winnipeg corpus contains 22 recordings from 6 Canadian English-speaking families with 5- to 20-month-old infants (McDivitt & Soderstrom, 2016; Soderstrom et al., 2018). Finally, the San Joaquin Valley Corpus contains 40 recordings from 15 US English-speaking families from Central California with infant ages ranging from 3 to 18 months (Warlaumont et al., 2016). The ACLEW dataset is a subset of these corpora, consisting of 10 recordings selected from each corpus to maximize diversity in socioeconomic status (SES) and age. For each of the 10 recordings, random uniform sampling was used to select a total of fifteen 2-minute segments for manual annotation while ensuring that the segments did not overlap (300 minutes per corpus) (Soderstrom et al., 2021). Given the diversity of participants, corpora, and uninformed random sampling, we assume that the annotated data captures a typical variation in longform data from English-speaking families. The annotations include utterance timestamps, utterance transcripts, and speaker identities, with all four corpora annotated according to the same protocol using the ACLEW annotation scheme[3].

In the experiments, the basic unit of processing and evaluation is one manually annotated utterance, whose manual transcription is used as a reference for ASR WER calculation. To obtain adult speech, we extracted utterances associated with female and male adult speakers, as labelled in the manual annotation. We excluded utterances shorter than 300 ms from our experiments, because early testing revealed ASR to be highly unreliable for very short utterances. These utterances accounted for approximately 5% of all utterances, consisting mostly of backchannels such as “yeah” or “okay”. The resulting dataset consists of 193 minutes of annotated audio and a total of 8,570 utterances.

[2] https://sites.google.com/view/aclewdid/home
[3] https://osf.io/b2jep/

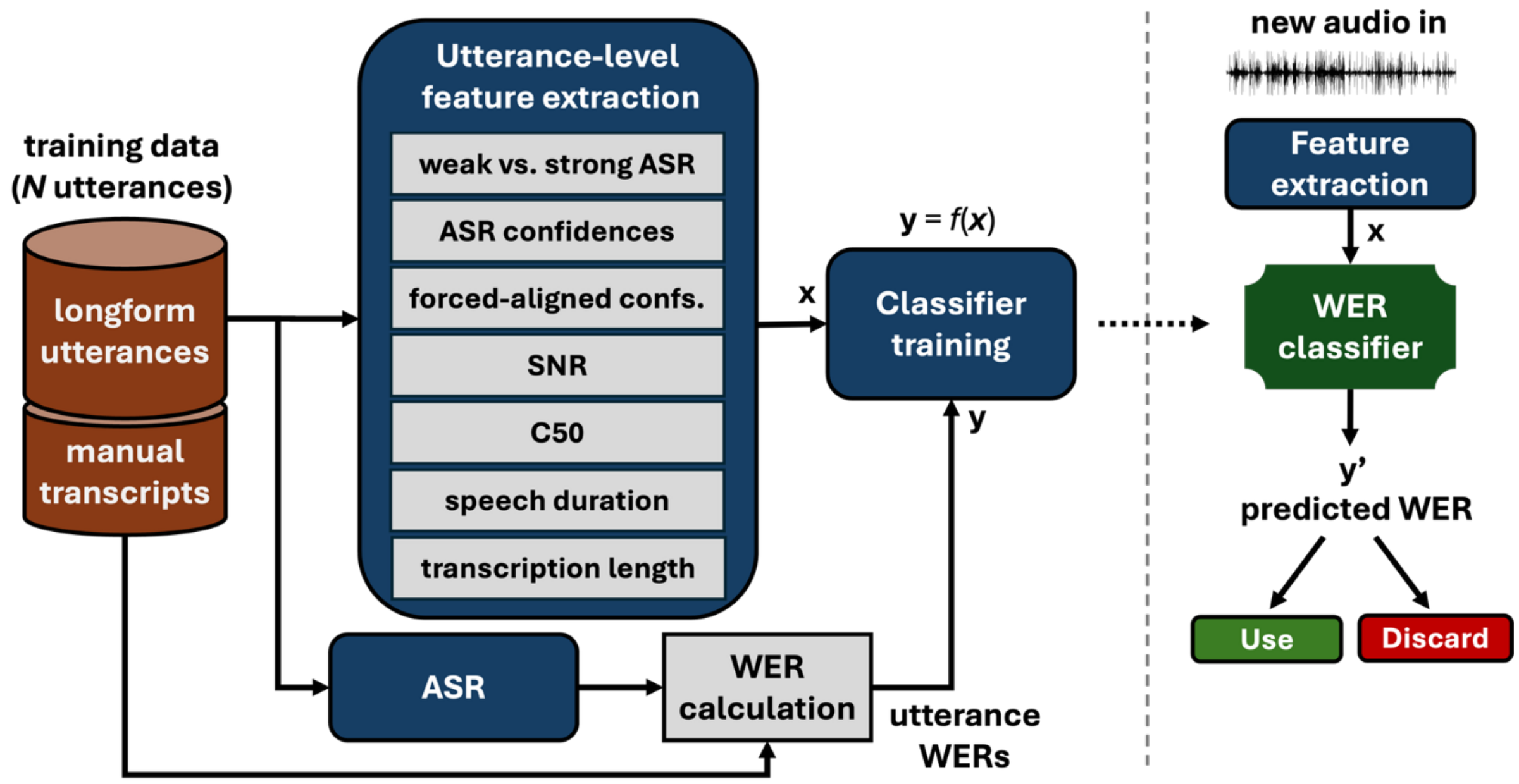

**Figure 1.** A schematic view of the proposed pipeline for estimating the reliability of ASR transcripts on child-centered longform audio data.

### 2.2 An overview of the pipeline

Our pipeline for automatic ASR sample selection is based on an ensemble of classifiers that is trained to predict the estimated utterance-level WER of an ASR system used for speech transcription. The approach relates to ensuring the quality of speech transcription using ensembles of ASR models (e.g., Gitman et al., 2023; Bhogale et al., 2024) and confidence-based filtering of ASR outputs (e.g., Naderi et al., 2024), but transforms the techniques into a functioning solution for the longform audio context.

Classification of expected WER is based on a number of features that can be calculated for arbitrary unlabeled audio. The tested list of features includes differences in outputs of a weak and strong ASR system, ASR and forced-alignment confidence scores, and automatic signal quality estimates. During training, the classifier is trained with samples with known WERs from a predetermined ASR system (here: Whisper large) and tuned for a target precision/recall trade-off, which then enables WER estimation on previously unseen data. Fig. 1 presents an overview of the pipeline, and individual components are detailed in the following subsections.

More specifically, one feature subset is based on applying *small* and *large* Whisper models as weak and strong ASR models to automatically recognize the speech content of each utterance (Radford et al., 2023). The difference between the outputs of the weak and strong ASR systems is measured similarly to WER, calculated using normalized Levenshtein distance between the word strings (transcriptions) from the two systems. Levenshtein distance corresponds to the number of edits required to make the two strings match, thereby accounting for word insertions, deletions, and substitutions. Confidence scores of the Whisper small and large models correspond to the posterior probabilities assigned by the models to their outputs, which we summarize across the variable-length outputs using the mean, minimum, and maximum log-probability across the recognized words in each utterance. Word-level confidence scores were obtained using the *whisper-timestamped* implementation[4].

We additionally perform a forced alignment between the automatically obtained word transcriptions and the acoustic speech signal with word-level confidence scores. Alignment confidence was quantified as the mean, minimum, and maximum log-probability of the aligned words within each utterance. Forced alignment is produced using the wav2vec2-base-960[5] model (Baevski et al., 2020) as implemented in *whisperX*[6] by Bain et al. (2023). To model the audio quality, estimates for speech-to-noise ratio (SNR) and C50 room reverberation score are also extracted using the Brouhaha toolkit[7] (Lavechin et al., 2023). As the final two features, duration of the utterance and number of words in the automatic transcription by the *large* Whisper model are included.

All the features are standardized with z-score normalization.

### 2.3 Classifiers and experimental setup

The prediction of ASR WER can be approached either as a statistical regression or classification task. While linear or non-linear regression from features to scalar WER values may appear more intuitive, our initial pilot testing indicated that the classification approach results in better overall performance. To this end, we defined two classes of utterances: those with a low WER ("acceptable" WER for transcription) and those with a high WER (utterances to "discard" from the transcription). Based on early pilot experiments with subsets of the data and using support vector classifiers (SVC; Cortes & Vapnik, 1995), we set the class threshold to 30% WER as a compromise between data balance in the two categories while keeping the WER as low as possible in the target category. On the current data, this corresponds to approximately 72 minutes (37%) of speech considered to have acceptable WER and 121 minutes (63%) considered not usable for recognition. In this context, recall as a performance metric indicates the proportion of the 72 minutes of speech that was successfully detected by the pipeline.

In our experiments, we compared several classifiers as well as their ensemble model using the feature sets described in Section 2.3 to distinguish between the two classes of utterances. The classifiers included logistic regression (Verhulst, 1845), a linear-kernel SVC, CatBoost (Prokhorenkova et al., 2018), LightGBM (Ke et al., 2017), and XGBoost (Chen & Guestrin, 2016). Each of these models produces a confidence score in range [0,1] for the low-WER class detection, a higher score indicating stronger confidence for the detection.

For the ensemble classifier, model predictions were combined by averaging the confidence scores of the involved classifiers. A preliminary examination of calibration curves was conducted, calculating Brier scores for all separate models. Since none of the models showed substantial miscalibration, we decided not to include additional calibration steps prior combining the models into the ensemble.

---

[4] https://github.com/linto-ai/whisper-timestamped
[5] https://github.com/facebookresearch/fairseq/tree/main/examples/wav2vec
[6] https://github.com/m-bain/whisperX
[7] https://github.com/marianne-m/brouhaha-vad

A cross-validation scenario with a leave-one-corpus-out protocol was used. In each validation fold, three speech corpora were used for a classifier training, and the remaining corpus was left for testing. The evaluation results reported below consist of precision and recall of low-WER class, as well as WER median and mean, which is the utterance-wise difference between the automatically produced and manual transcripts. Additionally, we evaluated AUC (area under the curve) to compare different classification setups.

The primary results are reported for a detection threshold of 0.6 which was found to provide a good balance of relatively high precision with moderate recall. More detailed analyses on the impact of the threshold choice are reported in Section 3.3.

### 2.4 Evaluation of a fully automatic speech transcription pipeline

Our primary experiments rely on manually annotated utterance boundaries and speaker identities. While useful for understanding performance of the ASR solution itself, such an approach does not reflect the use case in which no manual annotations are available—a common scenario in longform analysis and a motivation to use automated tools in the first place. To address this aspect, we also report results from a fully automatic pipeline for speech segmentation and speaker diarization, followed by our automatic selection of reliable transcripts. Speech segmentation and diarization were performed with the voice type classifier (VTC)[8] (Charlot et al., 2025; version 2.1), a tool which is specifically developed for processing child-centered longform recordings. VTC detects speech segments ("utterances") in an audio recording and assigns each of them to one of four speaker classes: adult female, adult male, key-child (with the microphone), and another child. In our experiments, we used speech segments assigned to the adult female and adult male classes. The VTC-extracted speech segments were automatically transcribed and classified as low or high WER using the same procedure as for manually segmented utterances, as described in Section 2.2.

The automatic pipeline can result in utterance boundaries that do not align with manually annotated utterances, sometimes also splitting long utterances into several shorter ones or merging several consecutive utterances into longer ones. In addition, the proposed filtering approach can leave some audio segments untranscribed. Consequently, standard WER calculation between hypothesized and manual transcripts is not possible at the utterance level and does not provide representative results at the level of the 2-minute clips. To address this, we developed a heuristic algorithm[9] to estimate the WER in situations where ASR hypotheses and reference transcripts can be asynchronous, can consist of different numbers of utterances for the same transcribed content, and where ASR may not attempt to transcribe all the speech in a long clip.

[8] https://github.com/LAAC-LSCP/VTC.
[9] https://github.com/SPEECHCOG/async_WER

The underlying principle of the algorithm is as follows: each automatically transcribed speech segment is compared with all reference transcription utterances that overlap with the segment, and WER is calculated separately for each of the possible pairings (in case of multiple overlapping candidate references) and stored in a list. In addition, WER is calculated separately after first concatenating two or more subsequent reference utterances into longer utterances, and potentially also concatenating two or more subsequent automatic segments. By testing all possible combinations of individual and merged segments that overlap in time between the automatic and reference transcripts, the pairing with the lowest WER is chosen, and the involved segments in both automatic and manual transcripts are removed from further analysis. The process repeats in a greedy manner from left to right, always finding the best pairing for the next unpaired automatic segment in time. If an automatic segment has no overlap with manual annotation, its WER is counted as to 100% (all words are insertions). The resulting WER can be viewed as the *precision* of the transcripts: what proportion of transcribed words were actually said in the data. In contrast, it does not reflect the *recall* of the system, i.e., what proportion of all manually annotated data were transcribed correctly. Recall must be separately measured as the proportion of manual transcripts that ended up paired with automatic segments.

The same cross-validation setup as described in Section 2.3 was used for the fully automatic pipeline, but manually segmented utterances of the three training corpora were always used to train the filter classifiers before filtering the automatic outputs on the test fold.

# 3 Results

## 3.1 Main results

Table 1 summarizes the evaluation results for the different classifiers. SVC, CatBoost, XGBoost, and LightGBM show very similar overall performance, with CatBoost and XGBoost achieving a slightly higher precision (by approximately 1 percentage unit). Logistic regression performs notably worse in terms of precision but results in a higher recall. However, its median WER within the low-WER class is above 0 (0.09), indicating poorer separation between the classes compared to the other approaches. The ensemble model provides a strong compromise, increasing recall with almost no loss in precision in comparison to the separate models. Note that the recall is calculated at utterance level, whereas different utterances may have different durations. This means that the proportion of all data transcribed does necessarily go hand-in-hand with the recall in the different compared setups.

Table 2 presents the detailed experimental results of a leave-one-corpus-out evaluation of the best model, i.e., ensemble of classifiers, where the subcorpora are abbreviated as follows: BER – SEEDLingS corpus, LUC – LuCiD Language 0–5, SOD – the McDivitt and Winnipeg corpora, and WAR – Warlaumont corpus. The quality of *low-WER* transcripts in terms of precision, recall, and mean WER significantly varies from one tested corpus to another. The recall in the BER case is higher than in the other cases, resulting in a lower mean WER than for the LUC and SOD. The precision for the WAR case is significantly higher than for the other cases which results in lower mean WER than for the other cases. The difference in the results may reflect the difference in the recording quality of the speech data from different corpora.

**Table 1.** Validation results for the different classifiers, where low-WER class is considered as the target (positive) class. The reported values correspond to averages across the four folds of the leave-one-corpus-out validation process. The overall proportion (%) of speech data classified as low-WER, and hence considered "reliably transcribed", is shown in the rightmost column.

| Classifier | AUC | | Selected utterances (low-WER class) | | | |
|---|---|---|---|---|---|---|
| | | Precision | Recall | WER median | WER mean | % of all speech transcribed |
| Logistic regression | 0.75 | 0.71 | 0.68 | 0.09 | 0.38 | 37 |
| SVC | 0.75 | 0.79 | 0.54 | 0.0 | 0.17 | 28 |
| CatBoost | 0.84 | 0.85 | 0.62 | 0.0 | 0.14 | 29 |
| LightGBM | 0.86 | 0.86 | 0.63 | 0.0 | 0.14 | 29 |
| XGBoost | 0.81 | 0.82 | 0.59 | 0.0 | 0.16 | 29 |
| Ensemble of classifiers | 0.82 | 0.81 | 0.61 | 0.0 | 0.16 | 30 |

**Table 2.** Leave-one-corpus-out cross validation results for automatic prediction of low-WER samples using the ensemble classifier. The last two columns show the baseline WER if no automatic sample selection is applied.

| Test corpus | Training corpora | Selected utterances (low-WER class) | | | | No selection | |
|---|---|---|---|---|---|---|---|
| | | Precision | Recall | WER median | WER mean | WER median | WER mean |
| BER | LUC, SOD, WAR | 0.81 | 0.67 | 0.0 | 0.16 | 0.35 | 0.44 |
| LUC | BER, SOD, WAR | 0.78 | 0.57 | 0.0 | 0.19 | 0.71 | 0.59 |
| SOD | BER, LUC, WAR | 0.82 | 0.58 | 0.0 | 0.16 | 0.61 | 0.55 |
| WAR | BER, LUC, SOD | 0.84 | 0.58 | 0.0 | 0.14 | 0.42 | 0.47 |

### 3.2 Manual error analysis

We also conducted a manual analysis of 200 utterances from the SEEDLingS corpus (BER) that had been predicted as belonging to the low-WER class but for which the Whisper transcription actually differed from the manual transcription. The analysis revealed that 66 utterances (33%) had a mismatch between the Whisper and manual transcriptions, but in which the automatic transcription could still be considered relatively accurate for many analytical purposes. Examples of such mismatches included:

- Colloquial vs. canonical spelling: the manual transcriptions contained contractions to reflect the colloquial word reduction in the spoken language (e.g., '*gonna*', '*alright*', '*kinda*') while the Whisper transcripts contained more canonical forms of language (e.g., '*going to*', '*all right*', '*kind of*').
- Difference in determiners: e.g. manual transcription with '*it's the cupcake*' was automatically transcribed as '*it's a cupcake*'.
- Resyllabification and mistranscription of vocalizations: e.g. manual transcription of '*ki ki ki ki ki ki*' and automatic transcription of '*Kiki, kiki, kiki*'.
- Isolated interjections with ambiguous spelling: e.g. manual transcription of '*yup*' versus automatic transcription '*yep*'. Note that these one-word utterances were scored with a WER of 1 even if only one character differed.
- Spelling of numerals: The manual transcriptions typically used spelled-out forms (e.g., '*twenty*'), whereas Whisper tended to transcribe numbers as numerals (e.g., *'20'*).

These error patterns highlight how automatic comparison of manual and ASR transcriptions significantly affects the estimates of automatic transcription quality if text normalization is not applied.

The analysis also revealed 84 utterances (42%) where the automatic transcription had inserted, substituted, or deleted words in comparison with the manual transcription to make the automatic transcriptions (e.g., *'that's what I'm thinking*', '*I mean us*', '*I just don't think it's fair*') more grammatically correct or complete than the manual ones (e.g., '[...] *what I'm thinking*', '*I mean a* [...]', '*I just don't think it's fair that* [...]'; brackets added to denote missing contiguity), where hesitations, restarts, and other properties of conversational language result in incomplete fragments of language when detached from the context. Moreover, manual transcriptions included some words that are not found in standard vocabularies of English, such as '*banoona*' or '*dax*'[10], which may be either misspellings or representations of special pronunciations used in the given CDS context.

The remaining 50 cases (25%) contained real mistranscriptions. Usually these were cases of partial mistranscription, with smaller or larger differences between the manual (e.g., '*there's nothing in here*', '*should hide it up on the couch*') and automatic transcriptions ('*there's nothing in there*', '*She eyed it up on the couch.*'). There were also certain cases of overlapping speech, for which the ASR system was not able to correctly transcribe both speakers. In such cases, the speech of the loudest speaker was typically transcribed. The closer the second speaker in loudness to the first, the more likely there were errors in the transcription due to the interfering speech from both speakers.

[10] The out-of-vocabulary words were detected using the NLTK toolkit (Loper & Bird, 2002).

### 3.3 Results on feature set and decision threshold optimization

We performed a series of feature selection experiments aimed at reducing the number of features involved in the classification (not reported in detail for conciseness). In short, these experiments revealed that the signal-quality features, SNR and C50, did not provide additional performance gains compared to the rest of the features. In contrast, exclusion of any of the other features led to various degrees of degradation in classifier performance. Consequently, the publicly available version of the ASR pipeline does not use SNR and C50 features for identifying the reliable samples for transcription.

We also analyzed the influence of various detection thresholds on the performance of the classifier ensemble, reported in Table 3 and Figure 2. While the choice of appropriate precision/recall -tradeoff may depend on the exact use case, two thresholds of 0.6 and 0.7 stand out with zero median WER while still having recall of at least 50%. They can thereby be used as default choices in the absence of further preference.

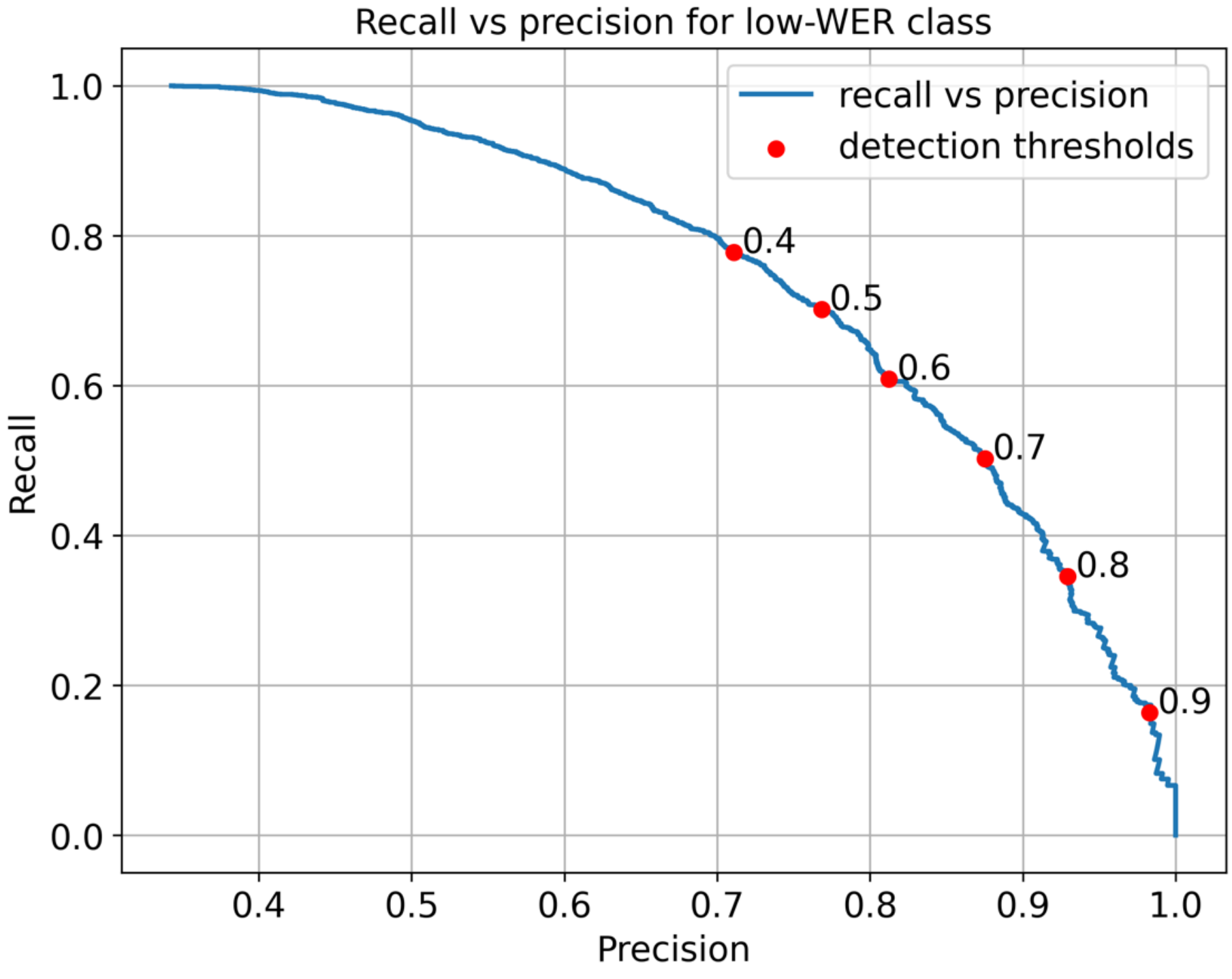

**Figure 2.** The relation between precision and recall of the ensemble of classifiers for low-WER samples across all the corpora. Values for various alternative detection thresholds are illustrated with red dots.

**Table 3.** The influence of the detection threshold on the resulting transcription quality (for the ensemble of classifiers). The presented values correspond to averages across all four folds in the leave-one-corpus-out validation process. Precision and recall are reported for the low-WER class. The overall

proportion (%) of speech data classified as low-WER, and hence considered "reliably transcribed", is shown in the rightmost column.

| Detection threshold | Precision | Recall | WER median | WER mean | % of all speech transcribed |
|---|---|---|---|---|---|
| no selection | — | — | 0.52 | 0.51 | 100 |
| 0.5 | 0.77 | 0.70 | 0.0 | 0.19 | 36 |
| 0.6 | 0.81 | 0.61 | 0.0 | 0.16 | 30 |
| 0.7 | 0.88 | 0.50 | 0.0 | 0.13 | 23 |
| 0.8 | 0.93 | 0.35 | 0.0 | 0.09 | 16 |
| 0.9 | 0.98 | 0.16 | 0.0 | 0.06 | 7.9 |

### 3.4 Results for the fully automatic speech transcription pipeline

Table 4 presents the results of a leave-one-corpus-out evaluation, when the ensemble of classifiers is applied to adult speech segments automatically extracted using VTC 2.1. Note that the WER median is calculated across the segment-specific WERs after the potential merging of consecutive utterances. The overall quality of automatic transcriptions is somewhat lower than in the manually segmented utterances because of additional errors introduced by the automatic diarization (see Charlot et al., 2025, for a performance analysis). Nevertheless, the improvement in transcription quality due to automatic selection is substantial, turning basically unusable transcripts into those with the median WER ranging from 0.03 to 0.17, depending on the corpus.

**Table 4.** Leave-one-corpus-out cross validation results for a fully automatic speech segmentation, diarization and transcription pipeline using the ensemble classifier. The last two columns show the baseline WER if no automatic sample selection is applied.

| Test corpus | Training corpora | Selected utterances (low-WER class) | | | No selection | |
|---|---|---|---|---|---|---|
| | | WER median | WER mean | % of all speech transcribed | WER median | WER mean |
| BER | LUC, SOD, WAR | 0.07 | 0.18 | 22 | 0.50 | 0.45 |
| LUC | BER, SOD, WAR | 0.17 | 0.26 | 25 | 0.67 | 0.56 |
| SOD | BER, LUC, WAR | 0.14 | 0.33 | 32 | 0.80 | 0.52 |
| WAR | BER, LUC, SOD | 0.03 | 0.20 | 25 | 0.62 | 0.45 |

Closer analysis of cases with the largest difference between automatic and manual transcriptions revealed one source of 'errors' which were caused by our procedure of WER estimation, and which lowered the estimates of transcription quality. When a manually extracted utterance was segmented by VTC into several shorter utterances, only a subset of these segments could be classified as reliable. As a result, the automatic transcription sometimes represented only part of the original utterance, leading to a high WER despite the transcribed portion being largely accurate (e.g., *'coming up, are the snakes living in your walls?'* vs '*coming up, are there snakes living in your walls, the answer may surprise you, after the break.*' yielded a WER of 0.53). In contrast, the manually segmented utterance might be considered as reliably transcribed when processed in its entirety by the automatic filter. This limitation could not be fully addressed within the current evaluation framework. However, it also exemplifies cases in which the automatic pipeline deviates from a standard manual annotation process. Consequently, the reported WER values should be interpreted as a conservative estimate of transcription quality, but the interpretation depends on how the resulting transcripts are to be used for further analysis.

## 4. Lexical analysis of automatic and manual transcriptions

We also wanted to understand how usable the ASR output from our pipeline was for basic research purposes. To this end, we performed lexical analysis to investigate how accurately word frequencies derived from automatic transcriptions reflected the ground-truth frequencies derived from the manual annotations. For the frequency estimates, all the manually transcribed utterances in our dataset and corresponding automatic transcriptions produced by the *large* Whisper model were used. We performed automatic lemmatization and part-of-speech (POS) tagging of automatic and manual transcriptions and then compared the counts of words (lemmas) in both transcriptions. Manually segmented but automatically transcribed and selected utterances were used as the ASR outputs. The main research question was whether the automatic selection of well-recognized utterances improves the correlation between the word counts in the manual and automatic transcriptions.

The lemmatization and POS tagging were performed with Stanza toolkit (Qi et al., 2020) using UPOS tags from the Universal Dependencies framework, for which the previously reported POS tagging accuracy ranges from 96.20% up to 99.15% depending on the dataset[11]. When processing spoken speech, parsing and tagging quality usually degrades due to the lack of punctuation marks, presence of disfluencies, and ungrammaticality of some speech content. Hence, we expected POS tagging to introduce some noise into the analyses.

Statistical analysis was performed using Pearson correlation on $\log_{10}$-scaled word counts. All word counts were incremented by 1 prior to taking the logarithm to avoid taking the logarithm of zero. The analysis was performed separately for all the words and then for a set of main POS categories. Table 5 presents the comparison of experimental results for a set of selected automatic transcriptions of utterances in comparison with the results obtained when processing the entire speech dataset with no selection. All the presented results are statistically significant at $p < 0.001$, and hence we only report the correlation coefficients. The obtained results (Table 5) show that there is a notable increase in the correlation coefficient between our pipeline for sample selection and when applying ASR to all speech. This is even though the amount of speech is larger in the naive approach, which helps to average out random errors in estimates.

[11] https://stanfordnlp.github.io/stanza/performance.html

**Table 5.** Pearson correlation between automatic and manual log-scaled word counts for different POS categories without and with sample selection.

| Part of speech | No sample selection | Proposed pipeline |
| --- | --- | --- |
| all words | 0.85 | 0.94 |
| nouns | 0.64 | 0.81 |
| verbs | 0.88 | 0.94 |
| adjectives | 0.81 | 0.86 |
| adverbs | 0.92 | 0.97 |
| pronouns | 0.95 | 0.95 |

The analysis shows that the results vary across different parts-of-speech. Since verbs, adverbs, and pronouns were well transcribed even for very challenging inputs, the automatic selection of reliable transcriptions has little effect on the correlation coefficient. For nouns and adjectives, the improvement is much larger. The potential reason why POS categories associated with main content words (nouns, verbs, adjectives) have lower correlation than others is due to higher type-token ratio (TTR) and consequently lower lexeme-specific frequencies in these categories. For instance, the TTR of nouns and adjectives in the analyzed reference manual transcriptions were 0.34 and 0.22, respectively, whereas verbs had TTR of 0.11, adverbs 0.06, and pronouns 0.01. The more there are tokens per type in the POS category, the less impact individual estimation errors have on the correlation score. In addition, in cases of very noisy speech, the ASR system struggles to identify these words based on the acoustic cues, whereas pronouns tend to be more predictable in context (or at least come from a closed set of alternatives).

Fig. 3 illustrates the correlation between word counts in the manual and automatic transcriptions when processing utterances selected automatically as reliable ones and when processing all the utterances. The plots also illustrate that excluding rare words in the automatic transcriptions (frequency less than five) significantly improves the correlation of word counts in the transcriptions for both cases. In the case of the proposed sample selection pipeline, the correlation is almost perfect ($r = 0.99$) when the rare words are excluded.

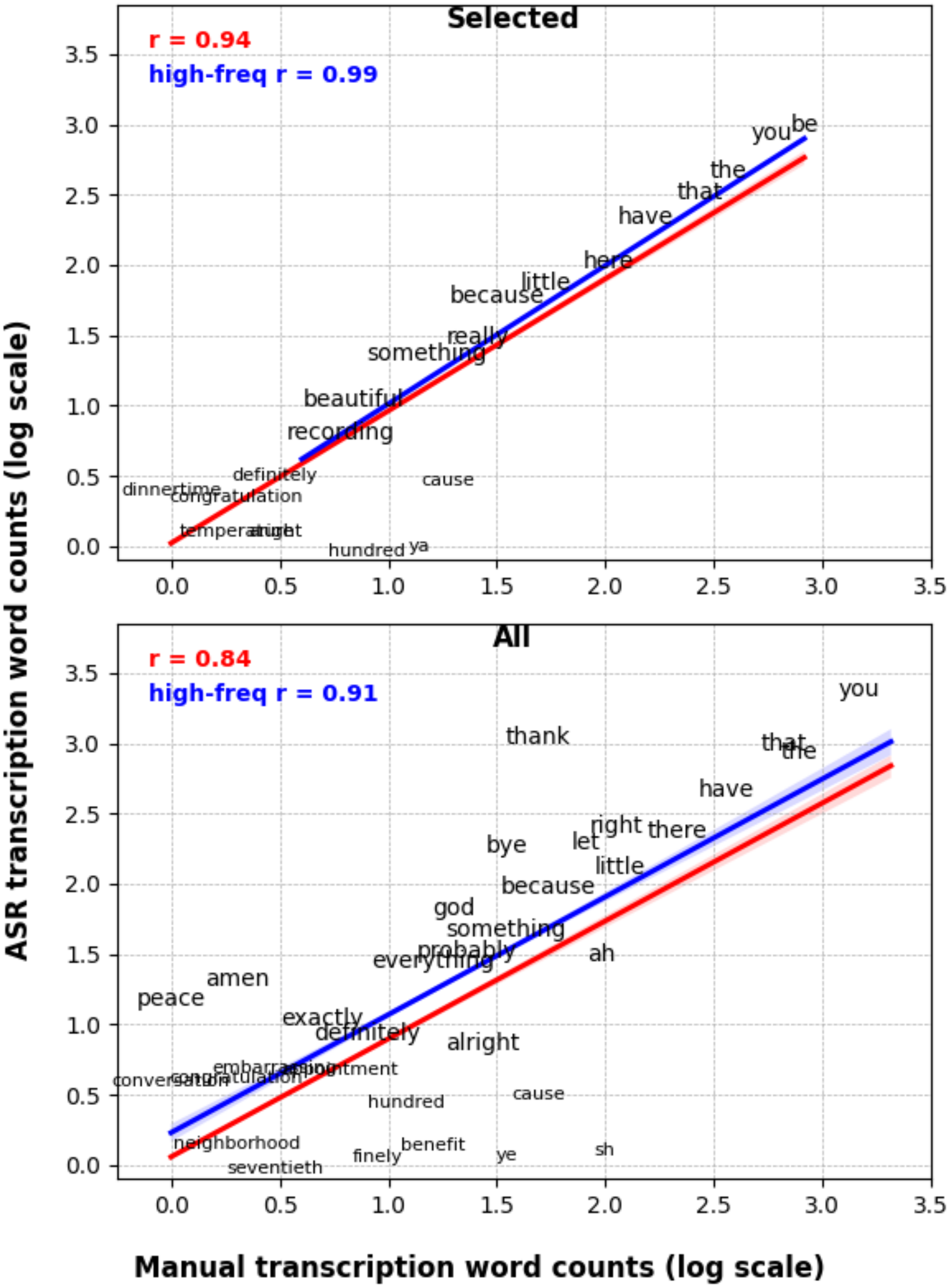

**Figure 3.** Log-scaled counts of words in automatic and manual transcriptions for utterances with (top) and without (bottom) automatic selection of reliable transcriptions. Pearson correlation was calculated between the log-scaled counts. The red line indicates the correlation for all words; the blue line shows the correlation for words that appear at least five times in the automatic transcriptions. Word labels are shown only for a small subset of words to improve visual clarity.

Fig. 4 illustrates the corresponding word frequency analyses for nouns and verbs, since they differed in terms of their overall transcription accuracy (Table 4). The visualization shows that the accuracy of noun transcription improves substantially when using automatic sample selection for ASR. The figures also highlight the text normalization issues for some words. For instance, the frequencies of ‘*cause*’ and ‘*dan*’ are underestimated while ‘*because*’ and ‘*dun*’ are overestimated by ASR (Fig. 3, bottom plot), although it is likely that these variants represent the same lexemes in the conversational data.

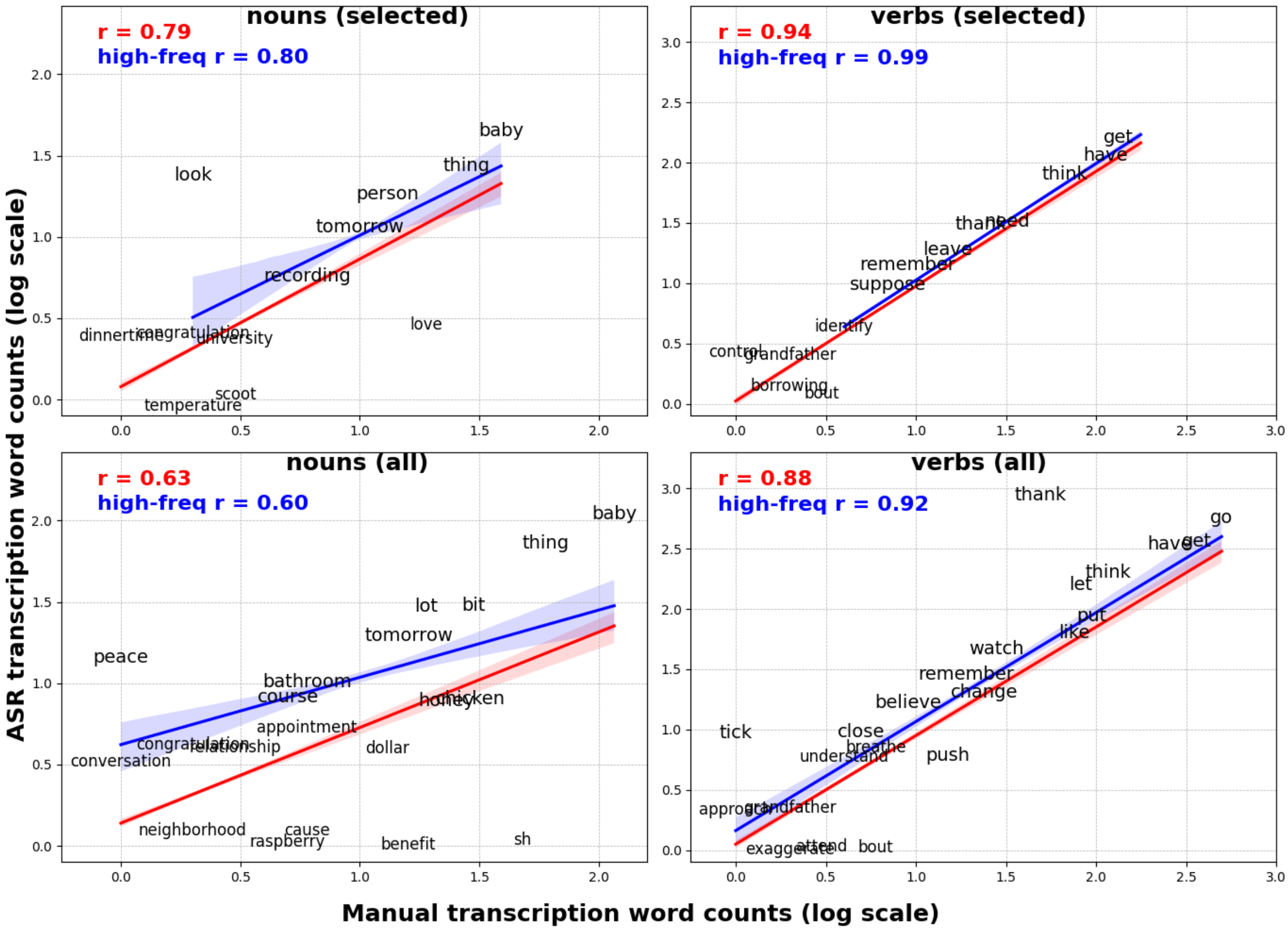

**Figure 4.** Log-scaled counts of *nouns* (left) and *verbs* (right) in automatic and manual transcriptions in *selected* (top) and *all* (bottom) utterances. Pearson correlation was calculated between the log-scaled counts. The red line indicates the correlation for all words; the blue line shows the correlation for words that appear at least five times in the automatic transcriptions. Word labels are shown only for a small subset of words to improve visual clarity.

## 5. Discussion

This paper introduced an approach to automatically identify those utterances from child-centered longform audio that can be reliably transcribed using an automatic ASR system. We tested the approach using manually transcribed segments from four longform corpora and trained our classifier to work with the Whisper-large ASR model by OpenAI. We conducted experiments applying two segmentation strategies: manual (original) segmentation and automatic segmentation using the VTC tool. The results indicate that the approach successfully reduces the average WER of automatic transcriptions of manually selected utterances to below 20% and that of the automatically selected utterances to below 30% while maintaining a reasonable yield of speech audio. Word frequency analyses also indicate that word frequency estimates from the transcriptions are relatively accurate, especially for words that appear multiple times in the automatic transcriptions. Overall, this highlights the promise of the approach for performing ASR on child-centered recordings, enabling several linguistic analyses of children's language experiences that would otherwise be difficult—or at least very time-consuming—to conduct through manual annotation.

Manual analysis of incorrectly recognized transcripts with high confidence scores revealed that manual annotations are usually more reliable in cases of overlapping speech or when the utterance is not fully intelligible. In the case of overlapping speech, the ASR system often transcribes only the louder speaker, whereas human annotators typically transcribe both speakers correctly. Human annotators can also use contextual cues to correctly interpret short utterances that may be unintelligible when heard or processed in isolation.

Manual analysis of mismatches between manual transcripts and automatic ones selected as acceptable also revealed that both transcriptions should be normalized to canonical form before the calculation of WER. Additionally, the greatest source of errors was utterances that contained only onomatopoeia or isolated interjections. Thus, care should be taken when interpreting ASR errors for assessing transcript usability, as the WER metric may or may not reflect features that are relevant to the research question.

It is also important to note that ASR systems often make minor morphological, syntactic, and stylistic adjustments to transcriptions by correcting reduced word forms and by omitting or inserting words to better match expectations of the language model. While such behavior is useful in many everyday speech technology applications (e.g., human–machine interaction), it is suboptimal for developmental studies where it is important to document the exact word forms and syntactic constructions used by caregivers and/or children. Unfortunately, there is no straightforward solution for forcing an ASR system to output exact pronunciations while also maintaining high robustness to noise and other forms of acoustic variability in spoken language.

In addition, given the limited amount of annotated longform speech data, it is difficult to reliably estimate the accuracy of the ASR pipeline across a broader lexicon utilized by caregivers when producing child-directed language. Our analyses show that correlations of lexeme counts improve when the most infrequent words are discarded from the analysis. However, given that our data consists of only 30 minutes of audio per recording and 300 minutes per corpus, many of the words considered "rare" in the current analyses may correspond to mid-frequency items and hence become more reliable across more data. Still, the problem of rare words is likely to persist, not least since ASR systems might substitute rare words with more frequent (thus, more likely) words in poor signal conditions. This is especially the case in unexpected contexts, since reliance on language priors increases with acoustic uncertainty in machine learning models, as in human listeners.

### 5.1 Limitations and future directions

One major limitation of this work is that we used only English data in our experiments. This means that the approach should be verified on other languages—particularly those with lower resources and, consequently, weaker ASR systems. However, the core idea of training a classifier to detect segments that a given ASR system can reliably transcribe should generalize to any language where an ASR system is available, provided that a representative set of manually transcribed data is available for training.

The second key limitation is that the proposed pipeline focuses exclusively on improving adult speech recognition, despite longform child-centered recordings containing speech produced by both adults and children. This choice was made to isolate a single source of variability and thereby to enable a clearer interpretation of the proposed methodology, since state-of-the-art ASR systems are known to recognize child speech substantially less accurately than adult speech (Patel & Scharenborg, 2024; Singh et al., 2026). The approach is also a spiritual successor to the current tools in the field, where adult speaker word counts from LENA or ALICE or speech durations from VTC outputs are often used to estimate children's language exposure. Extending the approach to child speech recognition remains an important direction for future work. Notably, child speech ASR for longform data is not only a data selection problem but will require technological advances in the ASR technologies as well—a topic currently actively pursued in the field of speech technology.

Besides child speech, future developments in general ASR technologies could make the present filtering approach less important. Ideally, a future ASR system would perform substantially better for both adult and child speech in challenging conditions and would also automatically provide reliable confidence estimates for its outputs (which is not the case with the current technologies). However, even human annotators often struggle with longform transcription, marking notable proportions of data as unintelligible (or 'xxx' in the minCHAT/ACLEW annotation scheme). Therefore, it does not seem inevitable that a near-future ASR system would result in comprehensive high-quality transcription without at least some kind of screening for transcription quality. Moreover, Whisper, despite being already a 4-year-old tool at the time of writing, is still the state-of-the-art for many speech recognition tasks in adverse conditions, suggesting that ASR technologies are no longer developing as fast as some other fronts in AI.

There are also several avenues for further technical improvement of the proposed filtering approach. First, we treated utterances as independent speech segments without considering context, whereas in real-life scenarios, it may be more appropriate to transcribe speech turn-by-turn rather than utterance-by-utterance. Second, the current transcription selection and ASR pipeline can be viewed as a cascade of separate classifiers. This structure might be simplified by developing a single end-to-end neural network architecture that produces both the speech transcript and a confidence score directly from longform audio. Third, speech enhancement applied prior to ASR could potentially improve performance, although state-of-the-art ASR systems, such as Whisper, already compensate for many types of noise due to the diversity and robustness of their training data.

### 5.2 Summary and conclusions

This work proposed an approach for automatic transcription of longform child-centered audio data. Instead of trying to transcribe all the data, the approach focuses on identifying those parts of the audio where transcription is reliable. The trade-off is that only a subset of the total speech audio is transcribed. The current ensemble classifier is able to detect acceptable (low-WER) transcripts with a precision of 0.81 and a recall of 0.61 when the decision threshold set to 0.6. In this configuration, up to 30% of the original speech data is selected as reliably transcribed—equivalent to 18 minutes per hour of speech present in the recordings. Despite the number of transcription errors that remain after applying the pipeline, the resulting error rates are modest, and the corresponding transcripts should be useful for a range of research questions involving child-centered data.

Based on estimated daily speech exposure ranging from 32 to 540 minutes (derived from Coffey et al., 2024), our approach would allow researchers to automatically obtain approximately 1.25 hours of transcribed speech per day for typical recordings. While the majority of speech still remains untranscribed, the potential benefits should be analyzed in the context of typical manual annotation efforts possible for such data, as well as the sheer volume of audio available in longform (typically daylong) recordings. Manually transcribing child-centered audio can take 1–2 hours per one minute of recording—especially when speakers and addressees are also annotated (see Casillas, 2022). This means that transcribing 1.25 hours of audio from a single daylong recording may require nearly a week of manual effort, without considering the scale of datasets consisting of hundreds of such recordings. Therefore, even a high precision but moderate recall automatic transcription pipeline can yield quantities of adult speech transcripts that far exceed what is typically feasible with manual annotation resources.

In conclusion, we believe that the proposed solution will assist researchers in studying children's language experiences from longform recordings—particularly when the limitations of the approach are carefully considered during interpretation. At the same time, researchers must still rely on manual transcription for segments that are difficult to process automatically and may need to verify ASR output in cases where precise linguistic detail is critical.

## Acknowledgements

Daniil Kocharov was supported by L-SCALE project funded by the Kone Foundation and LINKAGE-SPEECH project funded by Research Council of Finland. Okko Räsänen was supported by ERC grant

EveryContext (grant no. 101229121). Program code for the reported classifier and lexical analysis of transcriptions is publicly available at https://github.com/SPEECHCOG/longform-ASR-pilot.

## Declarations

### Funding

The research was supported by the L-SCALE project funded by the Kone Foundation, LINKAGE-SPEECH project (grant no. 369379) funded by the Research Council of Finland, and by ERC grant EveryContext (grant no. 101229121). Views and opinions expressed are however those of the author(s) only and do not necessarily reflect those of the European Union or the European Research Council Executive Agency. Neither the European Union nor the granting authority can be held responsible for them.

### Conflicts of interest

The authors have no conflicts of interest to declare.

### Ethics approval

Not applicable, as no new data was collected.

### Consent to participate

Not applicable, as the study did not involve new data collection.

### Consent for publication

Not applicable, as the study did not involve new data collection and the data was used according to the original consents obtained by data owners. Manuscript does not contain identifiable information of original study participants.

### Availability of data and materials

Anonymized (tabular) lexical data have been deposited in GitHub https://github.com/SPEECHCOG/longform-ASR-pilot. The speech corpora used in this study consist of existing datasets collected earlier in various institutions. Three of the corpora are available through HomeBank, and access to the LuCiD Language 0–5 corpus can be requested from the data owner:

Bergelson, E. (2017). Bergelson SEEDLingS HomeBank corpus. doi:10.21415/T5PK6D.

McDivitt, K., & Soderstrom, M. (2016). McDivitt HomeBank corpus. doi:10.21415/T5KK6G.

Warlaumont, A.S., Pretzer, G.M., Mendoza, S. & Walle, E.A. (2016). Warlaumont HomeBank corpus. doi:10.21415/T54S3C.

Rowland, C.F., Bidgood, A., Durrant, S., Peter, M., & Pine, J.M. (2018). The Language 0-5 Project. University of Liverpool, 10.17605/OSF.IO/KAU5F.
The study was not preregistered.

**Code availability**

All program code used in the data preparation, experiments and result analyses are publicly available at https://github.com/SPEECHCOG/longform-ASR-pilot.

**Author contributions**

O.R. was responsible funding acquisition. O.R. and D.K. were responsible for the original conceptualization of the research and supervision of the research. O.R and D.K. participated in study design, development, experiments, analysis, and wrote and revised the manuscript together. O.R. and D.K. were responsible for initial implementations of the processing pipeline, and D.K. for linguistic analysis of transcripts. A.G. was responsible for the final implementation and optimization of the processing pipeline.